\documentclass[10pt,letterpaper,twocolumn]{article}

\usepackage{ol2}
\usepackage{epstopdf}
\usepackage{amsmath,amssymb}
\usepackage{times}
\usepackage{color}
\usepackage[font=small]{caption}
\begin{document}

\twocolumn[ 

\title{Efficiency of dispersive wave generation from a dual-frequency beat signal}

\author{K. E. Webb,$^{1,*}$ M. Erkintalo,$^{1}$ Y. Q. Xu,$^{2}$ G. Genty,$^{3}$ and S. G. Murdoch$^{1}$}

\address{$^1$Physics Department, University of Auckland, Private Bag 92019, Auckland 1142, New Zealand \\
$^2$Department of Electrical and Electronic Engineering, The University of Hong Kong, Pokfulam Road, Hong Kong, China.\\
$^3$Tampere University of Technology, Optics Laboratory, FI-33101 Tampere, Finland\\
$^*$Corresponding author: kweb034@aucklanduni.ac.nz
}

\begin{abstract}
The emission of dispersive waves (DWs) by temporal solitons can be described as a cascaded four-wave mixing process triggered by a pair of monochromatic continuous waves (CWs). We report experimental and numerical results demonstrating that the efficiency of this process is strongly and non-trivially affected by the frequency detuning of the CW pump lasers. We explain our results by showing that individual cycles of the input dual-frequency beat signal can evolve as higher-order solitons whose temporal compression and soliton fission govern the DW efficiency. Analytical predictions based on the detuning dependence of the soliton order are shown to be in excellent agreement with experimental and numerical observations.

\end{abstract}

 ] 

The emission of dispersive waves (DWs) by ultrashort pulses is a central process in nonlinear fiber optics. It plays a key role in the generation of short wavelength components in fiber supercontinua~\cite{husakou_supercontinuum_2001,cristiani_dispersive_2004, dudley_supercontinuum_2006, skryabin_colloquium:_2010, travers_blue_2010}, and can be harnessed to produce tunable laser sources in the visible, mid-IR, and even in the deep-UV ~\cite{chang_highly_2010,liu_all-fiber_2012, joly_bright_2011,travers_ultrafast_2011,mak_tunable_2013}. DW generation has traditionally been described in the context of soliton propagation~\cite{wai_nonlinear_1986,akhmediev_cherenkov_1995}, but the underlying physics is in fact more general, and equivalent radiation has been reported even from pulses propagating in the normal dispersion regime~\cite{genty_effect_2004,erkintalo_cascaded_2012, webb_generalized_2013,conforti_dispersive_2013,conforti_resonant_2014}.

DW emission has only recently been described in the frequency domain, with cascaded four-wave mixing (CFWM) identified as the underlying mechanism~\cite{erkintalo_cascaded_2012}, and this picture has also been extended to describe event horizon dynamics of soliton interactions~\cite{EHarxiv14}. Specifically, it has been shown that CFWM, triggered by two monochromatic continuous waves (CWs), can be phase-matched via higher-order dispersion leading to the amplification of an idler whose frequency coincides with that of the DWs emitted by the time-domain beat signal~\cite{erkintalo_cascaded_2012,EHarxiv14}. A significant conclusion of these studies was that, in the undepleted pump approximation, the phase-matched frequency does not depend on the frequency spacing of the CW pumps: the number of steps in the cascade changes, but the DW frequency remains the same. However, this analysis did not consider in detail whether the pump spacing impacts on the \emph{efficiency} of the process. For very low cascade orders and negligible pump depletion the problem can be solved analytically~\cite{erkintalo_cascaded_2012}, but this approach becomes intractable as the number of significant cascade components increases. Earlier experiments on CFWM-assisted pulse train generation have shown that temporal compression of the time-domain beat signal can be influenced by the frequency spacing~\cite{trillo_nonlinear_1994, pitois_generation_2002, fatome_20-ghz--1-thz_2006}, and even soliton fission has been predicted from modulated quasi-CW fields ~\cite{kutz_enhanced_2005, genty_route_2009, kelleher_fission_2012}. As both processes are known to result in efficient DW generation~\cite{cristiani_dispersive_2004, travers_ultrafast_2011}, these studies hint that phase-matched CFWM could also depend on the pump spacing. Yet, so far no study has been presented.

In this Letter, we report extensive experiments and numerical simulations that uncover a rich and unexpectedly complex set of CFWM dynamics as the pump frequency separation is varied. We explain our results theoretically by interpreting individual cycles of the input beat signal as solitons whose order depends on the pump frequency spacing. Significantly, we show that efficient DW amplification occurs due to the higher-order soliton compression of the time-domain beat signal. Our experiments confirm the manifestation of higher-order soliton dynamics within modulated quasi-CW fields~\cite{genty_route_2009, kelleher_fission_2012}, and could impact on the design of applications relying on CFWM~\cite{pitois_generation_2002, fatome_20-ghz--1-thz_2006}.

We consider two CW pumps $\omega_{\pm\text{p}}$ propagating in an optical fiber. Their four-wave interaction results in the formation of a frequency comb whose spacing is set by the pump detuning $\Delta = \omega_{+\text{p}}-\omega_{-\text{p}}$~\cite{mckinstrie_four-wave-mixing_2006, cerqueira_highly_2008}. As detailed in~\cite{erkintalo_cascaded_2012}, the $n^{\text{th}}$ component of the comb can be amplified even if none of the elementary four-wave mixing (FWM) processes are phase-matched. Assuming a cubic dispersion profile and negligible pump depletion, this can be achieved for any cascade order $n$ by choosing the pump detuning $\Delta_n=-3\beta_2/\left[\beta_3\left(n+1/2\right)\right]$, where $\beta_2$ and $\beta_3$ are the second- and third-order dispersion coefficients at the pump centre frequency $\omega_\text{c} =(\omega_{+\text{p}}+\omega_{-\text{p}})/2$, respectively. The frequency of the resulting phase-matched component is then $\omega_\text{DW}=\omega_\text{c} + \left(n+1/2\right)\Delta_n=\omega_\text{c}-3\beta_2/\beta_3$, and corresponds to the frequency of a DW emitted by a soliton centred at $\omega_\mathrm{c}$~\cite{erkintalo_cascaded_2012}. This analysis reveals the phase-matched frequency $\omega_\text{DW}$ to be independent of both $n$ and $\Delta$, but it does not provide information on the efficiency of the frequency conversion process.

\begin{figure}[t]
\centerline{\includegraphics[clip=true,width=8.3cm]{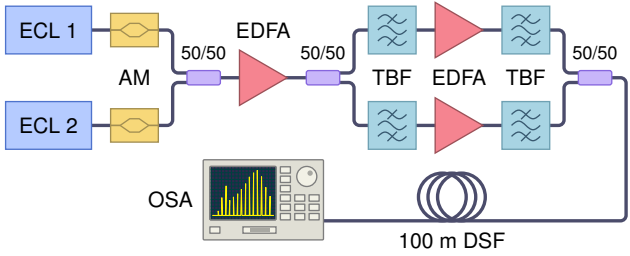}}
\caption{Experimental setup.}
\label{schematic}
\end{figure}

In order to study the impact of the pump frequency spacing, we use the experimental setup in Fig.~\ref{schematic}. We derive a dual-frequency field from two tunable telecommunications-band external cavity lasers (ECLs) that are amplitude modulated (AM) with a 40:1 duty cycle to produce 1.6 ns square pulses. The pumps are combined with a 50:50 coupler, and amplified with a low-noise L-band erbium-doped amplifier (EDFA), before being split and re-amplified using separate high-power L-band EDFAs. The pumps are filtered before and after the second pair of EDFAs with 0.4~nm tunable bandpass filters (TBFs) to remove amplified spontaneous emission. The two pumps are then recombined with a 50:50 coupler, such that their peak powers are 5~W and their polarizations are parallel aligned. The combined dual-frequency field is then launched into 100~m of dispersion-shifted telecommunications fiber (Corning DSF), and the output spectrum is recorded using an optical spectrum analyzer (OSA). By maintaining a fixed centre wavelength (1580.46~nm) while changing the frequency detuning $\Delta$ of the two ECLs, we are able to excite a wide range of cascade orders. Specifically, for each order $n$, the laser detuning $\Delta$ is set to a theoretically estimated value $\Delta_n$ that is obtained using the full phase-matching condition given in~\cite{erkintalo_cascaded_2012}, with dispersion coefficients up to fourth order ($\beta_2=-2.7$~ps$^2$km$^{-1}$, $\beta_3=+0.14$~ps$^3$km$^{-1}$, $\beta_4=-7\times10^{-4}$~ps$^4$km$^{-1}$). We then quantify the DW conversion efficiency as $P_n/P_\mathrm{tot}$ where $P_n$ is the power at the $n^{\text{th}}$ component and $P_\mathrm{tot}$ is the total integrated power.

\begin{figure}[t]
\centerline{\includegraphics[width=8.3cm,clip]{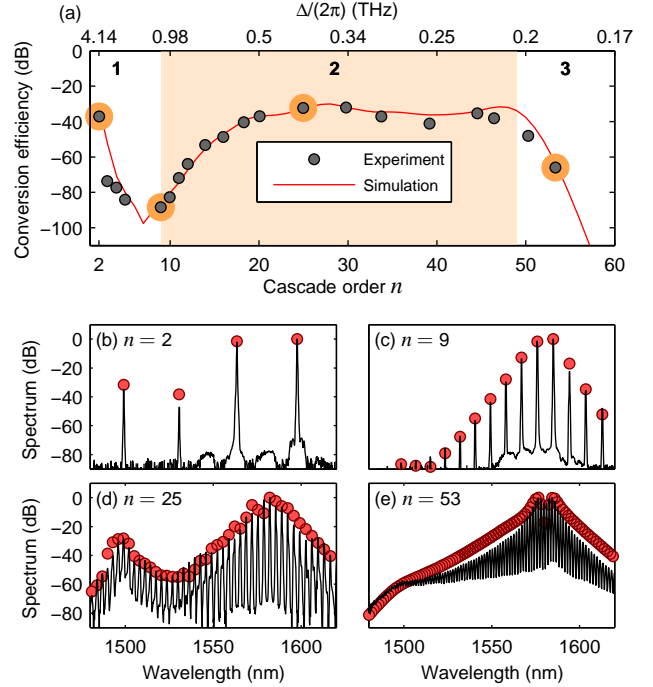}}
\caption{(a) Numerically simulated (solid curve) and experimentally measured (circles) CFWM conversion efficiency as a function of cascade order $n$ (top: pump frequency detuning $\Delta$). Highlighted circles correspond to spectra shown in (b-e). (b-e) Experimental (solid curve) and numerically simulated (circles) output spectra for cascade orders (b) $n=2$, (c) $n=9$, (d) $n=25$, and (e) $n=53$.}
\label{expr1}
\vskip -4mm
\end{figure}

Our experimental results are summarized in Fig.~\ref{expr1}. Specifically, Fig.~\ref{expr1}(a) shows the conversion efficiency as a function of cascade order (top axis: pump detuning), whilst \mbox{Figs.~\ref{expr1}(b-e)} show, respectively, the output spectra for selected pump detunings (and corresponding $n^{\text{th}}$ order cascades): $\Delta/2\pi=4.11$~THz ($n\approx2$), $\Delta/2\pi=1.09$~THz ($n\approx9$), $\Delta/2\pi=0.41$~THz ($n\approx25$), and $\Delta/2\pi~=~0.19$~THz ($n\approx53$). Also shown are results from numerical simulations using a generalized nonlinear Schr\"{o}dinger equation [red solid line in Fig.~\ref{expr1}(a); red filled circles in Fig.~\ref{expr1}(b-e)]. Our simulations use the experimental parameters specified above, and the nonlinear coefficient $\gamma=2.5$~W$^{-1}$km$^{-1}$. Stimulated Raman scattering is included for completeness.

As can clearly be seen from Fig.~\ref{expr1}, the pump frequency spacing substantially impacts on the DW conversion efficiency and output spectral characteristics. We can identify three distinct regimes [highlighted in Fig.~\ref{expr1}(a)]: (\textbf{1}) For low cascade orders ($n\leq9$) the conversion efficiency decreases monotonously; (\textbf{2}) for intermediate orders ($9<n<49$) the efficiency recovers and settles to a quasi-constant value; (\textbf{3}) for very large cascade orders ($n>49$) the efficiency again diminishes and eventually falls to zero. This behaviour is captured by experiments and simulations alike, and the two are in excellent qualitative and quantitative agreement.

We can describe these results qualitatively in the frequency domain by considering three competing processes that contribute to translating energy to the phase-matched sideband. First, cascaded nonlinearities typically diminish with order~\cite{durfee_phase_2002}, which explains the initial fall of the conversion efficiency in~Fig.~\ref{expr1}(a) [region \textbf{1}]. However, as the cascade order increases the pump frequency detuning decreases, resulting in smaller phase-mismatches between the elementary FWM processes. This leads to larger elementary coherence lengths, which allows for the intermediate sidebands to experience longer power law amplification, and to more efficiently convey energy towards the phase-matched sideband. Moreover, for sufficiently small pump detunings the first generated sidebands will experience \emph{exponential parametric} amplification, further enhancing the growth of all subsequent sidebands (recall that the cascade is iterative in that the $n^{\text{th}}$ sideband is driven by the $(n-1)^{\text{th}}$ sideband). Finally, the number of distinct cascade paths that result in the generation of photons at the phase-matched sideband increases as the overall number of sidebands increases, enhancing the overall efficiency~\cite{durfee_phase_2002}.

\begin{figure}[t]
\centerline{\includegraphics[clip=true,width=8.3cm]{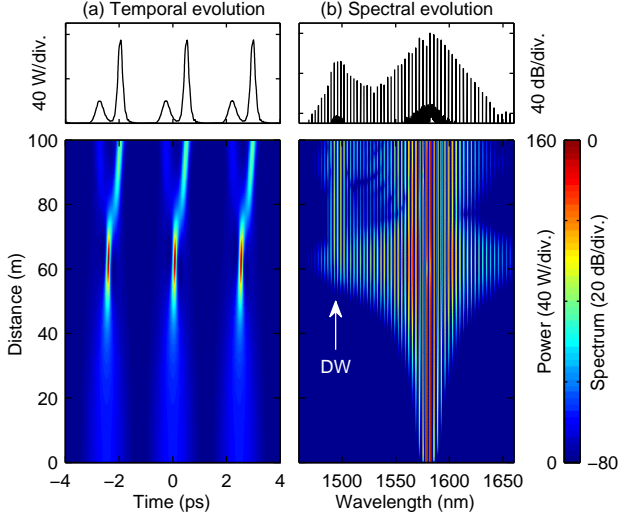}}
\caption{Results from numerical simulations showing (a) temporal and (b) spectral evolution of an incident dual-frequency beat signal with frequency spacing $\Delta/(2\pi)=405$~GHz and corresponding cascade order $n=25$. Top curves show the output profiles in more detail.}
\label{sim}
\end{figure}

For large $n$, the significant role of a very large number of sidebands, compounded by substantial pump depletion [see for example Fig.~\ref{expr1}(e)], renders quantitative frequency-domain analysis intractable. To gain further insight, we therefore investigate the time-domain dynamics in more detail. To this end, Fig.~\ref{sim}(a) and (b) show the temporal and spectral evolution corresponding to the output spectrum in Fig.~\ref{expr1}(d), respectively. We see how individual cycles of the initial sinusoidal beat note experience temporal compression, accompanied by the generation of multiple FWM components~\cite{trillo_nonlinear_1994}. As can be seen, the phase-matched components in the vicinity of $\omega_\mathrm{DW}$ are generated precisely at the point where the time domain signal exhibits maximal compression. Upon further evolution, the temporally compressed beat note can be seen to split into two constituents.

\begin{figure}[h]
\centerline{\includegraphics[clip=true,width=8.3cm]{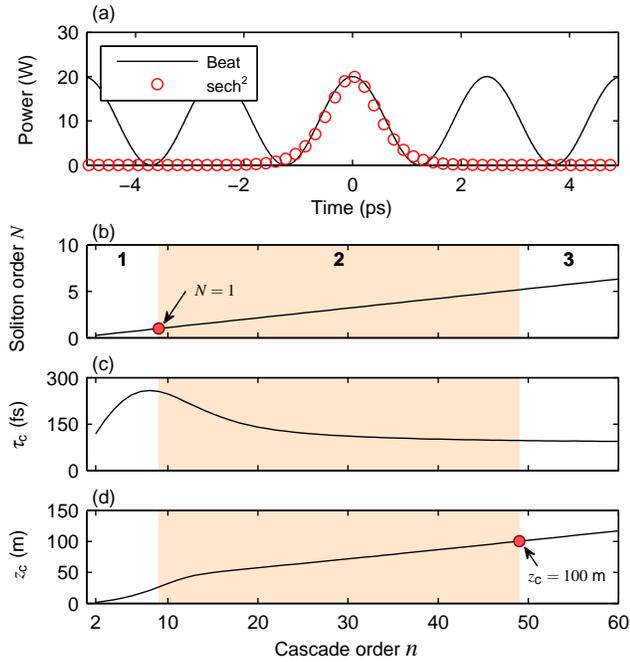}}
\caption{(a) Comparison of a dual-frequency beat signal and a $\mathrm{sech}^2$ pulse. (b) Calculated soliton order, (c) simulated FWHM duration of the maximally compressed beat cycle, and (d) the fiber length at which the maximum compression occurs as a function of cascade order. Regions \textbf{1-3} are the same as in Fig.\ref{expr1}, and can now be seen to be defined as $N<1$, $N>1$, and $z_\text{c}>100~\text{m}$, respectively.}
 \label{sol1}
 \vskip -10pt
\end{figure}

Figure~\ref{sim} shows that individual cycles of the input beat signal evolve quasi-independently, each displaying dynamics characteristic of higher-order solitons: strong temporal compression precedes fission into fundamental solitons~\cite{dudley_supercontinuum_2006}. Similar sub-pulse dominated fission dynamics has been previously reported in the context of supercontinuum generation excited by coherently and incoherently modulated quasi-CW fields~\cite{kutz_enhanced_2005, genty_route_2009, kelleher_fission_2012}. To gain more insights, we adapt the approach of Ref.~\cite{genty_route_2009}, and approximate an individual cycle of the input beat signal $P(t) = 4P_\mathrm{cw}\cos^2\left(\Delta/2 \cdot t\right)$ as a hyperbolic secant pulse with peak power $P_\mathrm{s} = 4P_\mathrm{cw}$ and duration $T_0 = \pi/(2\Delta)$. (Note that the duration $T_0$ is chosen such that the energy contained within a single modulation cycle equals the energy of the $\mathrm{sech}^2$ pulse~\cite{travers_blue_2010, kutz_enhanced_2005, kelleher_fission_2012}.) Figure~\ref{sol1}(a) illustrates the quality of this approximation using the parameters in Fig.~\ref{sim}, and we see fair agreement. We can therefore anticipate the cycles of the input beat signal to evolve as solitons whose orders are given by:
\begin{align}
N=\sqrt{\frac{\gamma P_{\text{cw}}}{\lvert\beta_2\rvert}}\frac{\pi}{\Delta}.
\label{solorder}
\end{align}
Recalling that the cascade order $n$ is inversely proportional to $\Delta$, Eq.~\eqref{solorder} shows that the soliton order $N$ increases linearly with $n$, as is shown in Fig.~\ref{sol1}(b) for our experimental parameters. Here we again highlight the same three regions as in Fig.~\ref{expr1}(a), and note that for the specific parameters in Fig.~\ref{sim} the soliton order $N\approx2.7$, which is in excellent agreement with the observed higher-order soliton dynamics.

We are now in a position to more substantially explain the conversion efficiency dynamics observed in our experiments and simulations [Fig.~\ref{expr1}(a)]. We first note that the pump spacing (cascade order) for which the individual cycles of the beat signal fulfill the fundamental soliton condition ($N=1$) is $\Delta/(2\pi) \approx 1.09~\mathrm{THz}$ ($n\approx9$). This point corresponds to the minimum efficiency in Fig.~\ref{expr1}. For larger cascade orders ($n>9$), the soliton order $N>1$, implying that the cycles of the beat signal will exhibit soliton pulse compression and subsequent fission. Indeed, in Fig.~\ref{sol1}(c) we explicitly show the FWHM at the point of maximum temporal compression, $\tau_\text{c}$, extracted from numerical simulations over the whole range of cascade orders examined in our experiment. We can clearly identify how efficient pulse compression takes place in Region \textbf{2}. The corresponding spectral broadening permits enhanced overlap with the phase-matched wavelength, resulting in efficient seeding of the dispersive wave ~\cite{cristiani_dispersive_2004, travers_ultrafast_2011}. This explains the increased efficiency for cascade orders $n>9$ [see Fig.~\ref{expr1}(a)], and is in line with the usual analytic description for DW emission\cite{akhmediev_cherenkov_1995}. We remark, however, that a quantitative comparison with this analysis is not straightforward since it assumes a fundamental soliton, which is not the case here. 

It is interesting to note that, for cascade orders $n>25$, the compressed duration $\tau_\mathrm{c}$ exhibits only small variations and approaches a constant value $\tau_\mathrm{c}\approx 94~\mathrm{fs}$ as $n\rightarrow\infty$, with the experimentally measured conversion efficiency in Fig.~\ref{expr1}(a) displaying a similar trend. This can be understood by observing that for higher-order solitons the compressed pulse duration can be estimated from an empirical formula (valid for large $N$)~\cite{agrawal_applications}: $\tau_\mathrm{c}\approx 1.763\cdot T_0/(4.1N)$. Within our approximation both $T_0$ and $N$ are inversely proportional to $\Delta$, yielding a constant compressed duration
\begin{equation}
\tau_\text{c}\approx0.22\sqrt{\frac{|\beta_2|}{\gamma P_{\text{cw}}}}\approx102~\text{fs},
\end{equation}
in good agreement with numerical observations. Finally, to explain the abrupt decrease in conversion efficiency for cascade orders beyond $n \approx 49$ [region \textbf{3}] we plot in Fig.~\ref{sol1}(d) the numerically simulated distance $z_\mathrm{c}$ at which maximal compression takes place. Significantly, we can see that for cascade orders $n>49$ the compression distance exceeds the 100 m fiber length available in our experiment. Thus, the experimental fiber length is too short to allow the beat cycle to achieve maximum temporal compression, explaining the decrease in conversion efficiency in this regime. This result is also amenable to approximative analytical description. Specifically, the distance at which a higher-order soliton is expected to fission can be coarsely estimated from $z_c\approx T_0^2/(|\beta_2|\cdot N)$~\cite{genty_route_2009, dudley_supercontinuum_2006}. In our approximation
\begin{equation}
z_\mathrm{c}=\frac{\pi}{4\Delta\sqrt{\gamma P_\mathrm{cw}|\beta_2|}}.
\label{zc}
\end{equation}
With our parameters, Eq.~\eqref{zc} predicts $z_\mathrm{c}$ will exceed the 100-m length of fiber used in our experiments for pump frequency detunings $\Delta/(2\pi) \lesssim 215~\mathrm{GHz}$ (cascade orders $n>46$). This is in good agreement with our observations, with the discrepancy arising from the intrinsic inaccuracy of the estimate.

The soliton analysis above is in remarkable agreement with the observed conversion efficiency in Fig.~\ref{expr1}(a), and fully explains the results for large $n$: In Region \textbf{2} the effective soliton order $N>1$, permitting fission and efficient DW seeding, whilst in Region \textbf{3} the compression distance $z_\mathrm{c}$ exceeds the available fiber length. Conversely, the measurements in Fig.~\ref{expr1} can be considered experimental evidence of fission and higher-order soliton compression from a modulated CW field.

There are several major conclusions from this work. We have experimentally and numerically shown that the pump frequency spacing significantly impacts on the efficiency of DW generation from a dual-frequency optical beat signal. We have explained our results theoretically in terms of higher-order soliton compression and fission of the input beat signal. Analytical estimates for the soliton characteristics are in excellent agreement with experimental observations. Our results therefore provide experimental support for the proposition that modulated quasi-CW fields can exhibit soliton fission dynamics~\cite{kutz_enhanced_2005,genty_route_2009, kelleher_fission_2012}. In addition, describing the evolution of a dual-frequency beat signal in terms of higher-order solitons could impact on the design of CFWM-assisted ultra-high repetition rate pulse train generation in fibers~\cite{pitois_generation_2002, fatome_20-ghz--1-thz_2006}, as well as in Kerr microresonators~\cite{herr_temporal_2013,coen_modeling_2013}. Finally, our results highlight that a dual-frequency beat signal constitutes a remarkably flexible and simple configuration for the experimental study of higher-order solitons and DW emission.

We acknowledge support from the Marsden Fund of the Royal Society of New Zealand, the University of Auckland Faculty Research Development Fund and the Academy of Finland (projects 130099 and 132279). We also thank J. M. Dudley for useful discussions.

\newpage

\textbf{Informational 5th page:}

\end{document}